\documentclass[aps,prl,twocolumn,a4paper,10pt, longbibliography]{revtex4-1}

\usepackage[T1]{fontenc}		
\usepackage[english]{babel}		
\usepackage[latin1]{inputenc}	
\usepackage{times}

\usepackage{amsmath}			
\usepackage{amssymb}			
\usepackage{bbm}				
\usepackage{mathtools}
\usepackage{gensymb}
\usepackage{simpler-wick}
\usetikzlibrary{tikzmark}
\DeclarePairedDelimiterX\Dirbraket[3]{\langle}{\rangle}%
{#1\,\delimsize\vert\,\mathopen{}#2\,\delimsize\vert\,\mathopen{}#3}

\usepackage[colorlinks=true, a4paper=true, pdfstartview=FitV,linkcolor=blue, citecolor=blue, urlcolor=blue]{hyperref}

\usepackage{graphicx}			
\usepackage{graphics}			
\DeclareGraphicsExtensions{.pdf}
\usepackage{color}		

\newcommand{\bea}{\begin{eqnarray}}
\newcommand{\eea}{\end{eqnarray}}

\newcommand{\bk}{\mathbf{k}}

\newcommand{\cg}{\textsl{g}}

\newcommand{\Tr}{\text{Tr}}

\begin{document}

\title{Integrating dynamical mean-field theory and diagrammatic Monte Carlo}

\author{Johan~Carlstr\"om }
\affiliation{Department of Physics, Stockholm University, 106 91 Stockholm, Sweden}
\date{\today}

\begin{abstract}
Dynamical mean-field theory (DMFT) is one of the most widely used theoretical methods for electronic structure calculations, providing self-consistent solutions even in low-temperature regimes, which are exact in the limit of infinite dimension. The principal limitation if this method is that it neglects spatial fluctuations, which become important in finite dimension. Diagrammatic Monte Carlo (diagMC), by contrast, provides results which are asymptotically exact for a convergent or resummable series, but are typically limited to high temperature as they depend on the analytic structure of the expansion. In this work, we present a framework for integrating these two methods  so that the diagrammatic expansion is conducted around the DMFT solution. This results in a series expansion conducted only in terms that explicitly depend on nonlocal correlations, and which is asymptotically exact. 
\end{abstract}
\maketitle

Strongly interacting fermions is both one of the most ubiquitous and also most challenging problems of theoretical physics. It arises in condensed matter systems, ultra-cold atomic gases, materials science, nuclear physics and in the interior of celestial bodies. Despite an intense effort and notable progress, theory fails to produce decisive results in many of these scenarios, and central questions remain open, some for decades. 

A major obstacle to theoretical progress in this field is the sign problem, which prevents the use of conventional quantum Monte Carlo techniques in fermionic systems. 
In response, a number of approximative methods have been proposed, ranging from DMFT \cite{PhysRevLett.62.324,PhysRevB.45.6479,RevModPhys.68.13} and generalizations thereof \cite{PhysRevB.75.045118,PhysRevB.77.033101,PhysRevLett.106.047004,PhysRevLett.110.216405,RevModPhys.77.1027,PhysRevLett.112.196402,PhysRevB.92.115109}, to wave function methods \cite{PhysRevLett.87.217002,PhysRevB.95.024506}, the density matrix renormalization group theory \cite{PhysRevLett.69.2863} and auxiliary-field quantum Monte Carlo \cite{PhysRevB.39.839,PhysRevLett.62.591,PhysRevB.40.506,afqmc}. Correlated systems are also extensively simulated with ultra-cold atomic gases \cite{PhysRevLett.115.260401,PhysRevLett.116.175301,nature14223}. Yet, despite considerable progress, a reliable phase diagram for even the most elementary fermionic many-body problem, namely the Hubbard model, has not been produced, and different numerical protocols produce results with notable discrepancies \cite{2006cond.mat.10710S,PhysRevX.5.041041}. 

A second obstacle to theoretical treatment of strongly correlated systems is the prospect of competing states that are situated closely in terms of free energy, making uncontrolled approximations potentially very misleading \cite{Dagotto257,2006cond.mat.10710S}. This fact points to the need for extremely accurate, preferably controlled, numerical methods which can be applied in the macroscopic limit. 

Diagrammatic Monte Carlo \cite{Van_Houcke_2010,PhysRevLett.119.045701,Rossi_2017} is a method which was developed specifically for unbiased treatment of many-body fermions, and is based on stochastic sampling of Feynman type graphs. Given a series that is either convergent or resummable, it provides asymptotically exact results directly in the macroscopic limit. 
The principal limitation of this method is that it relies on the analytic structure of the series. In metallic systems this typically prevents treating large expansion parameters, and also limits the applicability to fairly high temperatures. Thus far, near-zero temperature physics has only been reached in semimetals \cite{PhysRevLett.118.026403,PhysRevB.98.241102,PhysRevResearch.5.013069}.

The problems associated with a large expansion parameter means that a conventional expansion in the interaction part of the Hamiltonian is not viable for strongly correlated systems. This problem has now been partially overcome with the introduction of strong-coupling diagrammatic Monte Carlo  (SCDMC) \cite{PhysRevB.103.195147,0953-8984-29-38-385602,PhysRevB.97.075119,carlstrom2021spectral}, which relies on expansion in the nonlocal part of the Hamiltonian. Reliable results now exist for the Hubbard model in the strong coupling limit \cite{PhysRevB.103.195147}, the BCS-BEC crossover regime \cite{carlstrom2023disconnected} and magnetic moiré systems \cite{PhysRevResearch.4.043126}. The main issue at point is that the method is still limited to high temperatures.

An alternative path to diagrammatic simulation of challenging regimes is centered on the analytic structure of the series. Homotopic action relies on shifting the starting point of the expansion in order to improve convergence so that the point of interest in parameter space falls within the convergence radius \cite{kim2020homotopic}. By determining the analytical structure of a system in the weak coupling regime, it may be possible to reconstruct it in regimes where the series is not convergent \cite{PhysRevB.100.121102}.
By Borell resummation, it may even be possible to obtain results from a series with a zero convergence radius \cite{PhysRevLett.121.130405}. 
By themselves, these methods have provided access to $U/t\le 7$ in the Hubbard model. While this represent important progress, it is not by it self sufficient for addressing many key problems. It should be stressed however, that these methods are generally agnostic to the origin of the underlying series, and could for example be combined with alternative diagrammatic techniques.

In this work, we demonstrate that SCDMC can be integrated with DMFT to produce an asymptotically exact expansion around the mean-field solution, thus providing the means to systematically improve DMFT results. This is expected to dramatically expand the parameter regimes in which controllable results can be obtained, and shed new light on beyond-mean-field physics in correlated systems at low temperature. 

\subsection{Model}
DMFT is aimed at lattice fermions, and becomes exact in infinite dimension. In this regime, non-local interactions become trivial in that they only shift the chemical potential, and are correspondingly neglected. 
SCDMC is applicable to any lattice fermion problem, but was specifically developed to deal with strong contact interactions, by treating these non-perturbatively. 

In demonstrating how these methods can be integrated, we will rely on the Hubbard model, which takes the form
\bea
H=\hat{\mu}+\hat{t}+\hat{U},\; \hat{\mu}=-\mu \sum_{i\sigma} c_{i\sigma}^\dagger c_{i\sigma},\\
\hat{t}=\sum_{ij\sigma}t_{ij} c_{i\sigma}^\dagger c_{j\sigma},\;
\hat{U}=\sum_i U  c_{i\uparrow}^\dagger c_{i\uparrow} c_{i\downarrow}^\dagger c_{i\downarrow}.
\eea
However, it should be noted that long range interactions can be treated perturbatively with this approach \cite{PhysRevB.103.195147}.

\subsection{Dynamical mean-field theory}
\begin{figure}[!htb]
\includegraphics[width=\linewidth]{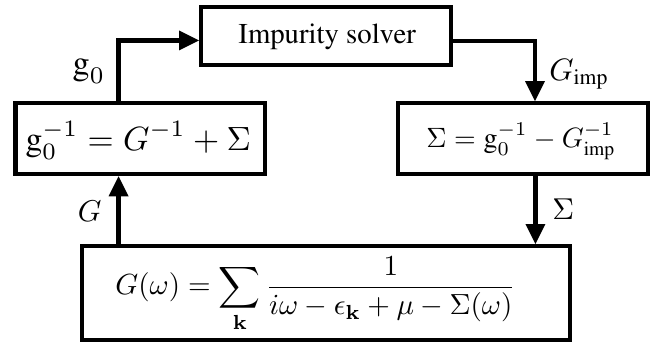}
\caption{
{\bf The DMFT cycle} is based on a self-consistency relation for the Greens function.  Solving the impurity problem with respect to an effective action $S_\text{eff}$, gives a local Greens function $G_\text{imp}(\omega)$. Comparison to the bilinear part of the effective action gives the self energy $\Sigma(\omega)$. Integrating over momenta provides a local Greens function $G(\omega)$, from which the bilinear part of the action, $\cg_0^{-1}$ is obtained and fed into the effective action. After convergence, a self-consistent solution for the Greens function is acquired. 
}
\label{DMFT}
\end{figure}

DMFT is based on mapping the many-body problem to an impurity model consisting of a single site embedded in a bath. Taking the limit of infinite dimension, the self energy becomes local, $\Sigma(\omega,\bk)\to \Sigma(\omega)$, which significantly simplifies the problem \cite{Vollhardt_2011}.
 The corresponding Greens function is then obtained by solving an impurity problem of the form
\bea
G_\text{imp}(\tau-\tau')=-\langle Tc(\tau)c^\dagger(\tau') \rangle_{S_{\text{eff}}},\label{impurity}
\eea 
where the effective action is given by 
\bea\nonumber
S_{\text{eff}}=-\int_0^\beta d\tau_1 d\tau_2 \sum_\sigma \bar{c}_\sigma(\tau_1) \cg_{0,\sigma}^{-1}(\tau_1-\tau_2)c_\sigma(\tau_2)
\\
+U\int d\tau n_\uparrow(\tau) n_\downarrow(\tau).\label{DMFTaction}
\eea
Since $\cg_0$ describes the bilinear part of the action, it is related to the full Greens function of the impurity problem by
\bea
\Sigma(\omega)=\cg_0^{-1}(\omega)-G_{\text{imp}}^{-1}(\omega).
\eea
Translation invariance implies that the self energy is the same for the impurity and the bath, giving
\bea
G(\omega)=\sum_\bk \frac{1}{i\omega -\epsilon_\bk +\mu -\Sigma(\omega)},\\
\cg_0^{-1}(\omega)=G^{-1}(\omega)+\Sigma(\omega).\label{g0}
\eea
From the equations \ref{impurity}-\ref{g0}, we obtain the DMFT cycle, as illustrated in Fig. \ref{DMFT}. The impurity problem, which is the only nontrivial step, can be solved by a range of different methods, including quantum Monte Carlo simulations and exact diagonalization \cite{RevModPhys.68.13}.

\subsection{Strong coupling diagrammatic Monte Carlo}

 \begin{figure*}[!htb]
 \hbox to \linewidth{ \hss
\includegraphics[width=\linewidth]{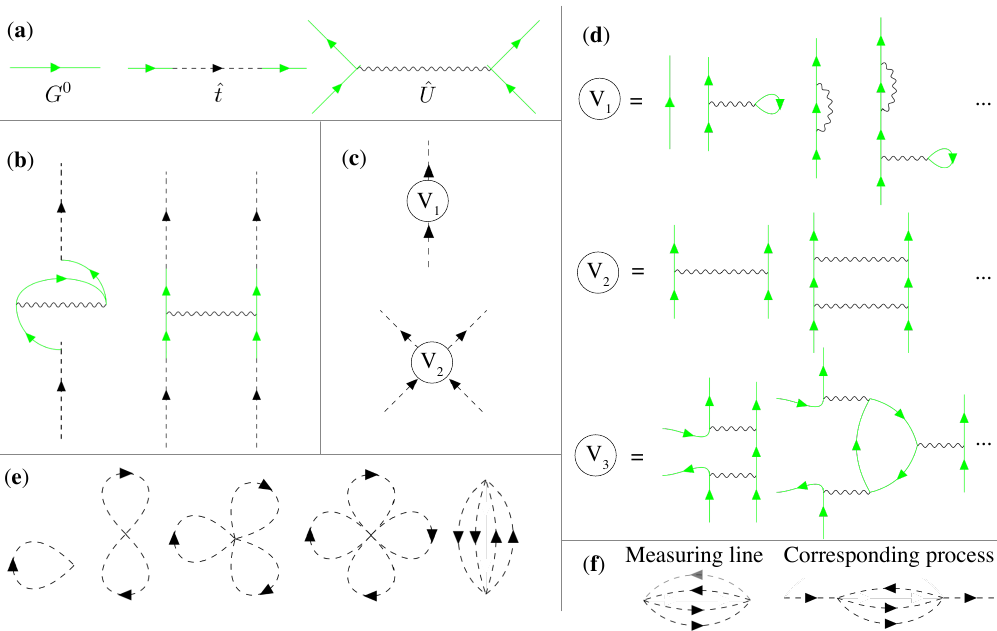}
 \hss}
\caption{
{\bf Overview of strong coupling diagrammatic Monte Carlo}. (a) Treating $\hat{t}$ and $\hat{U}$ as expansion parameters, we obtain two types of vertices. 
(b) The resulting diagrams consists of particles propagating along the $\sim t$ lines, which are then subject to corrections from local terms $\sim G^0$ and $\sim U$. These corrections can be classified as single-particle corrections and many-particle collisions respectively. (c) The set of all single particle corrections can be summed into a single particle vertex $\text{V}_1$, while the set of $N$-particle collisions can be sorted into vertices $\text{V}_N$. (d) The vertices are identical to the connected $N-$particle Greens functions in the atomic limit, making them an exactly solvable problem. 
(e) The expansion can then be conducted in vertices $\text{V}_N$ connected by the hopping integral. Using the dressed hopping integral, only skeleton graphs need to be sampled. (f) Observables are extracted using a measuring line, which is treated as an external line. The remaining part of the diagram provides a contribution to the polarization operator of the hopping integral, from which a dressed hopping integral and full Greens function can be obtained. 
}
\label{Overview}
\end{figure*}

SCDMC is based on expansion in the hopping integral $t$ and a set of vertices which describe scattering processes that are mediated by the contact interaction $\hat{U}$. The starting point for deriving this diagrammatic description is the treatment of nonlocal terms and interactions as a perturbation:
\bea
H_0=\hat{\mu} ,\; H_1=\hat{U}+\hat{t}.\label{H0H1}
\eea
Since $H_0$ is bilinear, eq. \ref{H0H1} can be treated via expansion and Wicks theorem:
\bea
\langle \hat{o}\rangle=\sum_n \frac{(-1)^n}{n!}\int_0^\beta d\tau_1... \langle T[H_1(\tau_1)...H_1(\tau_n)\hat{o}]\rangle_{H_o,c},\label{SCDMCexp}
\eea
where the subscript $c$ implies connected topologies. 
This gives two types of vertices corresponding to the expansion terms $\hat{t}$ and $\hat{U}$, as shown in Fig. \ref{Overview} (a). The resulting diagrams describe fermions dispersing according to $\sim t$ which are then subject to corrections that depend on the local terms $\hat{\mu}$ and $\hat{U}$, as exemplified in Fig. \ref{Overview} (b). The set of all such scattering processes for a given number of particles and corresponding quantum numbers (like spin), define a vertex $V_N$, where $N$ denotes the number of scattered particles, see Fig. \ref{Overview}, (c). Note that we suppress spin indices for brevity.   

Since $H_0$ contains no dispersion it follows that the bare Greens function is local,
\bea
G^0_{\alpha\beta}(i-j,\tau)=G^0_{\alpha\beta}(\tau)\delta_{i,j}. \label{locality}
\eea
The vertices $V_N$ can be expressed in terms of infinite sums as shown in Fig. \ref{Overview}, (d). These are equivalent to connected $N-$particle Greens functions obtained in the atomic limit, and can be written
\bea
V_N[\bar{O}]=\sum_n \frac{(-1)^n}{n!}\int d\bar{\tau} \langle  \hat{U}(\tau_1)...\bar{O}\rangle_{\hat{\mu},c},\label{VN}
\eea 
where $\bar{O}$ denotes the set of fermionic operators associated with the external lines, and the subscript $c$ implies connected topologies. Working in the atomic limit, the $N-$particle Greens function is an exactly solvable problem. The connected Greens function is then obtained by a recursion, where the disconnected parts are removed \cite{PhysRevB.103.195147}.

The vertices are connected by the hopping integrals $\sim t$ to form diagrams. The skeleton graphs up to order $4$ for the free energy are shown in Fig. \ref{Overview}, (e). Observables are obtained by inserting a measuring line into a diagram, as shown in Fig. \ref{Overview} (f): One line is tagged and treated as an external line. The remainder of the diagrams corresponds to an element of the polarization operator of the hopping integral, $\Pi$, which is the principal observable. The dressed hopping integral can then be obtained from a Bethe-Salpeter/Dyson type equation of the form
\bea
\tilde{t}(\omega,\bk)=\frac{1}{t^{-1}(\bk)-\Pi(\omega,\bk)}.\label{bethe}
\eea 
Expanding in $\tilde{t}$ and retaining only skeleton graphs (to avoid double counting) gives a self-consistent solution for the dressed hopping integral and the polarization operator. The Greens function can then be obtained from 
\bea
G(\omega,\bk)=\frac{1}{\Pi^{-1}(\omega,\bk)-t(\bk)},\label{SCTDMC-G}
\eea
from which it also follows that
\bea
\tilde{t}(\omega,\bk)=t(\bk)+t^2(\bk)G(\omega,\bk).\label{ttofG}
\eea
More complex observables can be obtained by using multiple measuring lines. 

\begin{figure}[!htb]
\includegraphics[width=\linewidth]{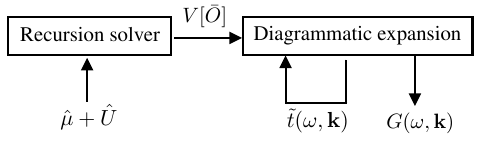}
\caption{
{\bf SCDMC cycle.} In the first stage, a data base of vertices are constructed from the local part of the Hamiltonian, $\hat{\mu}+\hat{U}$. In the second stage, expansion is conducted in these vertices and a dressed hopping integral $\tilde{t}$ to produce a new estimate for $\tilde{t}$, which is fed back into the diagrammatic expansion. The second stage is repeated until convergence is obtained. The Greens function is then obtained from $\tilde{t}$. 
}
\label{SCDMC}
\end{figure}

The SCDMC cycle is shown in Fig. \ref{SCDMC}: The first step consists of constructing all vertices up to a given order for all sets of operators $\bar{O}$ originating in the nonlocal terms $\hat{t}$. The second stage consists of expanding in $\tilde{t}$ to obtain $\Pi$, which in turn gives a new $\tilde{t}$. This process is repeated until convergence is obtained.

\subsection{Integration of dynamical mean-field theory and diagrammatic Monte Carlo}

To see how these two methods can be combined, we note that the impurity problem (Eq. \ref{impurity}) can be written as a series expansion:
\bea\nonumber 
G_{\text{imp}}(\tau-\tau')=Z_{H_1}^{-1}(\mu)\sum_n \frac{(-1)^n}{n!}\\
\times\int_0^\beta d\tau_i... \Tr e^{-\beta H_0} T[H_1(\tau_1,\tau_1')...c^\dagger(\tau')c(\tau)].\label{GimpExp}
\eea
Here, $H_0=\hat{\mu}$, $H_1(\tau,\tau')=\hat{U}\delta(\tau-\tau')+\Delta(\tau-\tau')c^\dagger(\tau')c(\tau)$. The partition function refers to contractions of $H_1$ evaluated with respect to $H_0$. Since $H_0$ is bilinear we can express this as a diagrammatic expansion 
\bea
G_{\text{imp}}\!=\!\sum_n \frac{(\!-\!1)^n}{n!}\!
\int_0^\beta d\bar{\tau}  \langle T[H_1(\bar{\tau}_1)...c^\dagger(\tau')c(\tau)]\rangle_{H_0,c},\;\;\;\;\;\label{GimpExpCon}
\eea
where the subscript $c$ denotes connected topologies and $\bar{\tau}_i=\{\tau_i,\tau_i'\}$. The term $\sim \Delta$ describes exchange of electrons with the bath \cite{Vollhardt_2011}, and takes the form
\bea
\Delta(\omega)=\frac{1}{\Big[\sum_\bk \tilde{t}(\omega,\bk)\Big]^{-1}+\Pi_{\text{imp}}(\omega)},
\eea
where $\Pi_{\text{imp}}$ denotes the polarization operator of $\Delta$ in the impurity problem.

  \begin{figure}[!htb]
\includegraphics[width=\linewidth]{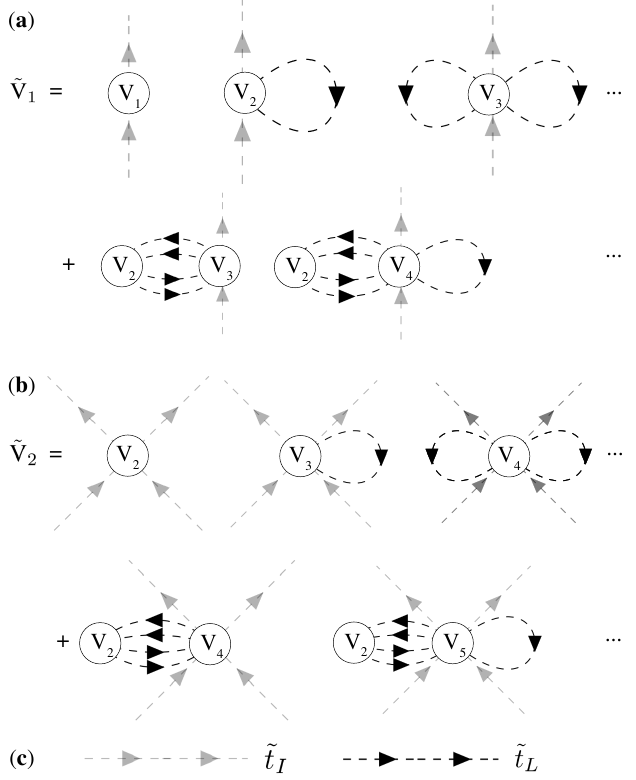}
\caption{
{\bf Mean-field like corrections.} (a) The single particle Greens function obtained from the impurity problem can be equated to a strong coupling expansion in the local part of the hopping integral, $\tilde{t}_L$, for a renormalized vertex $\tilde{\text{V}}_1$. (b) The $N$-particle connected Greens function of the impurity problem is equivalent to the infinite series for a corresponding vertex $\tilde{\text{V}}_N$. By integrating DMFT and SCDMC, these corrections can be accounted for nonperturbatively via an impurity solver, allowing for an expansion in the corrected vertices $\{\tilde{\text{V}}_N\}$. 
}
\label{DecNX}
\end{figure}

 In the next stage, we note that the expansion (\ref{GimpExpCon}) can be conducted in an exchange term which is dressed with respect to the impurity polarization operator, denoted $\tilde{\Delta}$, while retaining only the skeleton graphs (to avoid double counting). We then find 
\bea
\tilde{\Delta}(\omega)=\frac{1}{\Delta^{-1}(\omega)-\Pi_\text{imp}(\omega)}=\sum_\bk \tilde{t}(\omega,\bk).\label{BoldImp}
\eea
Separating the dressed hopping integral into local and inter-site parts
\bea
\tilde{t}_L(\omega)=\tilde{\Delta}(\omega),\; \tilde{t}_I(\omega,\bk)=\tilde{t}(\omega,\bk)-\tilde{t}_L(\omega),
\eea
the impurity problem can be cast into a strong-coupling expansion in $\tilde{t}_L$ for a renormalized single-particle vertex, as illustrated in \ref{DecNX}, (a). The vertex $\tilde{\text{V}}_1$ is equivalent to the impurity Grens function, and provides a contribution to the polarization operator of the full strong-coupling problem.  
Thus, given $\Delta$, the infinite series for $\tilde{\text{V}}_1$ can be accounted for non-perturbatively by solving the impurity problem.

This idea can be generalized by solving the impurity problem for connected $N$-particle Greens functions $G^N_{\text{imp},c}(\bar{\tau})$: Once again, we may cast this problem into a skeleton-graph expansion in the dressed exchange term $\tilde{\Delta}=\tilde{t}_L$, which produces an infinite series for a corrected $N$-particle vertex $\tilde{\text{V}}_N$, as shown in Fig. \ref{DecNX} (b). 
This problem is identical to the SCMDC expansion in the local part of the hopping integral $\tilde{t}_L$, implying that this infinite class of diagrams for a renormalized vertex may be replaced by the corresponding DMFT solutions.

In the next stage, we note that we can conduct the SCDMC expansion in $\tilde{\text{V}}_N$, as opposed to in the bare vertices. However, if we connect two vertices by the local hopping term $\tilde{t}_L$, we will produce an insertion that is by definition also an element of $\tilde{\text{V}}_N$, resulting in double counting. Consequently, when expanding in $\tilde{\text{V}}_N$, we are only allowed to connects vertices via the itinerant hopping $\tilde{t}_I$. From a diagrammatic point of view, this treatment is equivalent to including $\tilde{t}_L$ in the summation shown in Fig. \ref{Overview}, (d), to produce the vertices $\tilde{\text{V}}_N$ rather than their bare counterpart $\text{V}_N$, that only depend on $\hat{\mu}+\hat{U}$. At this point, $\tilde{t}_I$ becomes the expansion term. 

\begin{figure}[!htb]
\includegraphics[width=\linewidth]{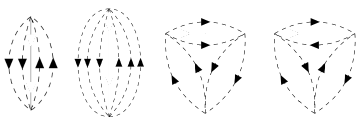}
\caption{
{\bf Topologies of the mean-field corrected series up to order 6.} In the corrected vertices $\{\tilde{\text{V}}_N\}$, the local part of the dressed hopping integral $\tilde{t}_L$ has effectively been integrated out. The corresponding expansion should therefore be conducted in the nonlocal part $\tilde{t}_I$, while retaining only skeleton graphs. As a result, there is only a small number of allowed topologies (4 graphs up to order N=6) that all explicitly depend on nonlocal correlations.  
}
\label{CEXP}
\end{figure}

The fact that the expansion term is non-local, and that we only permit skeleton graphs, leads to a sharp reduction of the number of diagrams that are permitted. Up to an expansion order 6, we only obtain 4 topologies, as shown in Fig. \ref{CEXP}. The remaining diagrams explicitly depend on nonlocal correlations, as all mean-field like contributions have been accounted for non-perturbatively.

\begin{figure}[!htb]
\includegraphics[width=\linewidth]{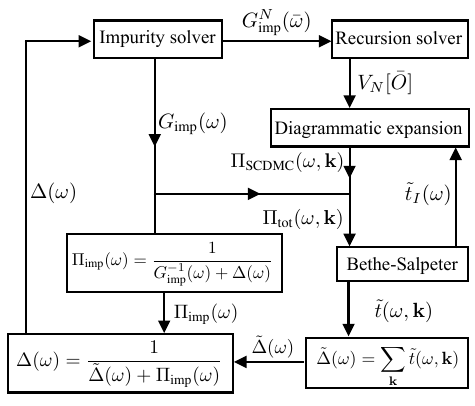}
\caption{
{\bf DMFT+SCDMC cycle.} The impurity solver allows mean-field like corrections to be integrated out, resulting in an expansion only in diagrams that explicitly depend on non-local correlations.  
}
\label{DMFT-SCDMC}
\end{figure}

With these results in place, we can proceed to construct an integrated DMFT-SCDMC algorithm in accordance with the diagram in Fig. \ref{DMFT-SCDMC}: 
\begin{enumerate}
	\item An impurity solver is used to obtain the single and many-particle Greens functions $G_\text{imp}^1,\;G_\text{imp}^N$. 
	\item The many-particle Greens function is fed into a recursive algorithm which removes its disconnected contributions, producing the renormalized vertices $\tilde{\text{V}}_N$ which are identical to connected Greens functions. 
	\item A diagrammatic expansion is conducted in the vertices $\tilde{\text{V}}_N$ and the itinerant part of the dressed hopping integral $\tilde{t}_I$, producing the beyond mean-field contribution to the polarization operator, $\Pi_\text{SCDMC}$.
	\item  $\Pi_\text{SCDMC}$ and $G_\text{imp}$ are combined, to form the full polarization operator $\Pi_{tot}$, which is fed into a Bethe-Salpeter/Dyson type equation (\ref{bethe}) to provide $\tilde{t}$ and $\tilde{\Delta}$. 
	\item $\Delta$ is obtained from $\tilde{\Delta}$ and $\Pi_{\text{imp}}$ and fed back into the impurity solver, while $\tilde{t}_I$ is inserted into the diagrammatic expansion. 
\end{enumerate}
The process above is repeated until convergence, providing a self-consistent solution for $\Pi$, which gives the dressed hopping integral, the Greens function and therefore also the spectrum of the system. Once convergence has been obtained, additional observables like spin-spin/spin-charge correlations can be obtained by using multiple measuring lines.

\subsection{Summary}
In conclusion, we have provided a framework for integrating DMFT with SCDMC which allows a systematic and asymptotically exact expansion around the mean-field solution. The remaining corrections in this series explicitly depend on non-local correlations, which now represent the expansion parameter, as all other contributions have been accounted for non-perturbatively. 

Since the renormalized vertices contain the same energy scales as the expansion parameter, it is very likely that convergence properties of the series is substantially improved compared to conventional diagrammatics, and this should give access to virtually exact results in parameter regimes where this was previously considered impossible. 
 
DMFT has an extremely wide applicability, ranging from condensed matter theory and ultra-cold atomic gases to material science. In the latter, it is often combined with DFT to conduct ab initio calculations of strongly correlated systems \cite{RevModPhys.78.865,GRANAS2012295}. Our method provides the means to dramatically improve results across these applications by systematically taking into account non-local correlation.

\bibliography{biblio.bib}

\end{document}